\documentclass{article}
\usepackage{spconf,amsmath,graphicx}
\usepackage{booktabs}
\usepackage{amsfonts}
\usepackage{multirow}
\usepackage{bbding}
\usepackage{cite}
\usepackage[urlcolor=blue]{hyperref}
\usepackage{subfigure}

\def\model{FullSubNet+}
\def\camodel{MulCA}

\title{\model{}: Channel Attention FullSubNet with Complex Spectrograms for Speech Enhancement}
\name{Jun Chen$^{1,\dagger}$\thanks{$^{\dagger}$ Work conducted when the first author was intern at Huya Inc.}, Zilin Wang$^1$, Deyi Tuo$^2$, Zhiyong Wu$^{1,3,*}$\thanks{$^{*}$ Corresponding author.}, Shiyin Kang$^2$, Helen Meng$^{1,3}$}
\address{
    $^1$ 
    Shenzhen International Graduate School, Tsinghua University, Shenzhen, China\\
    $^2$ Huya Inc., Guangzhou, China\\
    $^3$ 
         The Chinese University of Hong Kong, Hong Kong SAR, China\\
    \small{
        \{y-chen21, wangzl21\}@mails.tsinghua.edu.cn, 
        \{tuodeyi, kangshiyin\}@huya.com,
        \{zywu,hmmeng\}@se.cuhk.edu.hk
    }
}
\begin{document}
\ninept
\maketitle
\begin{abstract}
Previously proposed FullSubNet has achieved outstanding performance in Deep Noise Suppression (DNS) Challenge and attracted much attention.
However, it still encounters issues
such as input-output mismatch and coarse processing for frequency bands.
In this paper, 
we propose an extended single-channel real-time speech enhancement framework called \model{} with following significant improvements. 
First, we design a lightweight multi-scale time sensitive channel attention (\camodel{}) module which adopts multi-scale convolution and channel attention mechanism to
help the network focus on more discriminative frequency bands for noise reduction.
Then, to make full use of the phase information in noisy speech, our model takes all the magnitude, real and imaginary spectrograms as inputs. 
Moreover, by replacing the long short-term memory (LSTM) layers in original full-band model with stacked temporal convolutional network (TCN) blocks, we design a more efficient full-band module called full-band extractor. 
The experimental results in DNS Challenge dataset show the superior performance of our \model{}, which reaches the state-of-the-art (SOTA) performance and outperforms other
existing speech enhancement approaches.
\end{abstract}
\section{Introduction}
\label{sec:intro}

\begin{figure*}[!htb]
	\centering

	\subfigure[\model{} diagram]{\label{arch_a}\includegraphics[width=0.8\linewidth, height=0.12\linewidth]{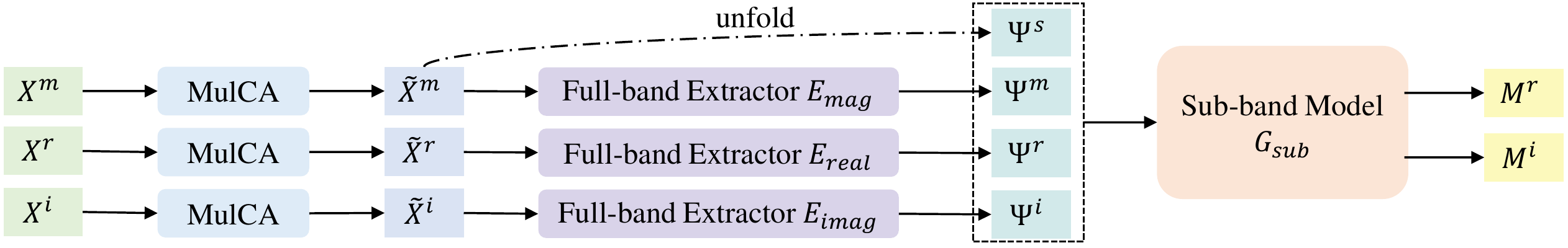}}
	\\
	\centering
	\subfigure[System flowchart on one branch of the model]{\label{arch_b}\includegraphics[width=0.82\linewidth, height=0.185\linewidth]{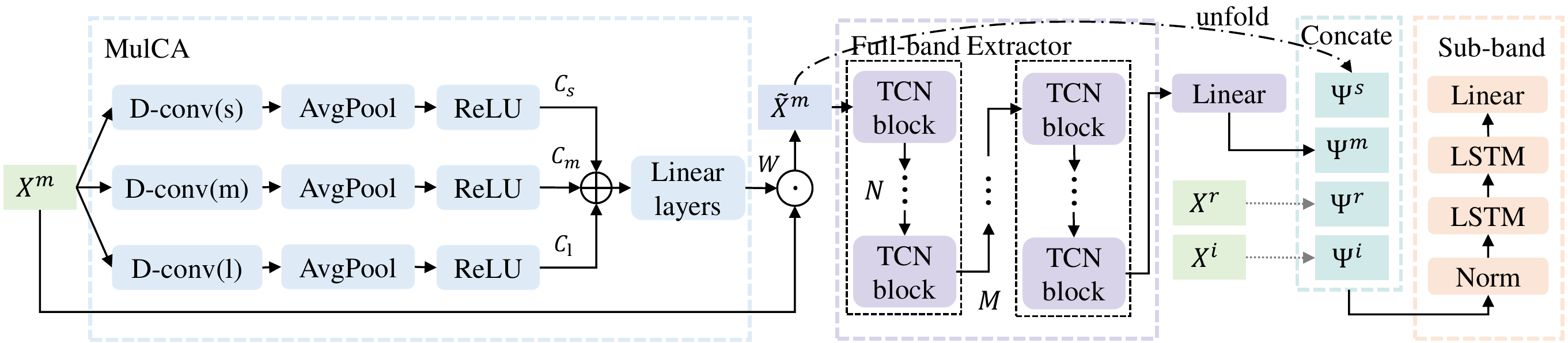}}
	\caption{(a) The overall diagram of the proposed \model{}. The model contains three branches consisting of an \camodel{} module and a Full-band Extractor, a concatenation operation, and a Sub-band Model.
	(b) The System flowchart on one branch of the model, which contains specific details of the \camodel{} module, Full-band Extractor, Concatenation operation, and Sub-band Model.}
	\label{fig:model_structure}
\end{figure*}

Single-channel (monaural) speech enhancement methods remove background noise from single-channel noisy audio signals, aiming to improve the quality and intelligibility of the speech.
Recently, deep learning-based methods have achieved promising results, especially in dealing with non-stationary noise under challenging conditions (e.g., low signal-to-noise ratio (SNR), reverberation).
Noisy speech can be enhanced by neural networks either in time-domain or frequency-domain.
The time-domain approaches \cite{2018wavenet, 2018waveunet, 2019tcnn} predict a clean speech waveform directly from the corresponding noisy speech waveform.
As another main-stream, the frequency-domain approaches \cite{2014regression, wang2018supervised, chen2017long, erdogan2015phase} generally take the noisy spectral feature as the input,
and their learning target is the clean spectral feature or the mask (e.g., Ideal Binary Mask\cite{ibm}, Ideal Ratio Mask\cite{irm}, complex Ideal Ratio Mask (cIRM)\cite{cirm}).
Overall, since the frequency domain signal has more distinguishable features for model learning \cite{yin2020phasen},
the frequency-domain approaches dominate the most SOTA speech enhancement methods.

FullSubNet \cite{hao2021fullsubnet} 
has been proposed
for single-channel frequency-domain speech enhancement. 
It consists of a full-band model and a sub-band model \cite{2020subband}, and performs joint optimization after connecting the two in series. 
The full-band model extracts the global spectral information and the long-distance cross-band dependencies. 
Subsequently, taking the concatenation of the sub-band units and the output of full-band model as the input, the sub-band model focuses on the local spectral pattern together with the extracted global information and models signal stationarity.
By means of this way, the FullSubNet can capture the global context \cite{2014regression, wang2018supervised} while retaining the  ability to model signal stationarity and attend the local spectral patterns.
As a result, FullSubNet achieves excellent results on DNS Challenge dataset \cite{dnschallenge}.

Despite the competitive performance, FullSubNet 
is still not perfect.
It has been pointed out that local patterns in the spectrogram are often different in each frequency band \cite{takahashi2017multi}.
However, in FullSubNet, the network takes the original whole frequency band as the input of full-band model, which degrades network's discriminative ability among different frequency bands in the input spectrogram.
Moreover, FullSubNet takes cIRM as its predicting target while being fed with only the magnitude spectrum which contains only magnitude information and lacks of the phase information. 
We argue in this work that the mismatching of input and output information is unreasonable, which will cause that the phase of enhanced speech deviates significantly with serious interference and limit the upper bound of performance.

In this paper, to address the above issues of FullSubNet, we propose a framework with channel attention on multiple input spectrograms, that is called \model{}.
We propose a lightweight \camodel{} module to give different attention degrees to different frequency bands of the input spectrogram.
At the same time, we design a full-band extractor composed of TCN blocks \cite{bai2018empirical} instead of the full-band model to reduce model size and save computation cost.
Furthermore, we deploy two additional independent full-band extractors to process the real and imaginary spectrograms to make full use of the phase information in noisy speech.
Experimental results show that \model{} prominently outperforms FullSubNet and other SOTA methods on the DNS challenge dataset,
which indicates the effectiveness of \camodel{} module and the use of noisy speech phase information for speech enhancement.

\section{Methodology}
\label{sec:format}

This paper focuses only on the denoising task in the short-time fourier transform (STFT) domain, and the target is to suppress noise and recover the speech signal (the reverberant signal received at the microphone). 
To accomplish this task, we propose a framework called \model{} with channel attention on multiple input spectrograms.
The model architecture is shown in Fig.\ref{arch_a}, which mainly consists of three main parts: three parallel \camodel{} modules, three parallel full-band extractors called $E_{mag}$, $E_{real}$, $E_{imag}$ respectively, and a sub-band model $G_{sub}$.


The proposed model takes the magnitude, real and imaginary spectrograms $\mathbf{X}^{m}$, $\mathbf{X}^{r}$, $\mathbf{X}^{i} \in \mathbb{R}^{F \times T}$ as inputs,
where $F$ and $T$ represent the total number of frequency bins and frames respectively.
The three parallel \camodel{} modules weight the above input spectrograms individually. 
Then the outputs of \camodel{} are fed to the next Full-band extractor and the weighted magnitude spectrogram $\widetilde{\mathbf{X}}^{m}$ will also be used to unfold for sub-band units $\mathbf{\Psi}^{s}$.
Next, the full-band extractors $E_{mag}$, $E_{real}$ and $E_{imag}$ extract the global spectral information from the weighted spectrograms, and then their outputs $\mathbf{\Psi}^{m}$, $\mathbf{\Psi}^{r}$ and $\mathbf{\Psi}^{i}$ are concatenated with the sub-band units $\mathbf{\Psi}^{s}$. 
Finally, taking the concatenation as input, the sub-band model $G_{sub}$ predicts the learning target cIRM $\mathbf{M}^{r}$ and $\mathbf{M}^{i}$.
In addition, the proposed model also supports output delay: to infer $M^{r}_{f, t-\tau}$, $M^{i}_{f, t-\tau}$, the future time steps are provided in the input sequence. In the following these modules are described in detail.

\subsection{Multi-scale Time Sensitive Channel Attention}
It has been pointed out in the previous work \cite{takahashi2017multi} that the local patterns in the spectrogram are often different in each frequency band: 
the lower frequency band tends to contain high energies, tonalities as well as long duration sounds, 
while the higher frequency band may have low energies, noise and rapidly decaying sounds.
Inspired by this, we regard different frequency bins as different channels, and give them different weights by using Channel Attention.
Specifically, we use a vector with dimension $F$ to represent the weight assigned to each frequency bin.
Then we perform a dot product operation on this weight vector with the spectrogram after a python broadcast to achieve the weighting:
\begin{equation}
\begin{split}
\mathbf{W} &= [W_0, \cdots, W_f, \cdots, W_{F-1}]^{T} \in \mathbb{R}^{F} ,\\
\widetilde{\mathbf{X}} &=  \mathbf{X} \odot  \mathbf{W}.
\end{split}
\end{equation}
where $\mathbf{W}$ represents the weight vector, and $\mathbf{X}$, $\widetilde{\mathbf{X}} \in \mathbb{R}^{F \times T}$ denote the spectrogram before and after weighting respectively. $\odot$ denotes the dot product.
Accordingly, the model will focus on frequency bands that play more significant roles in noise reduction.

\begin{figure}[!htb]
	\centering
	\includegraphics[width=0.3\linewidth]{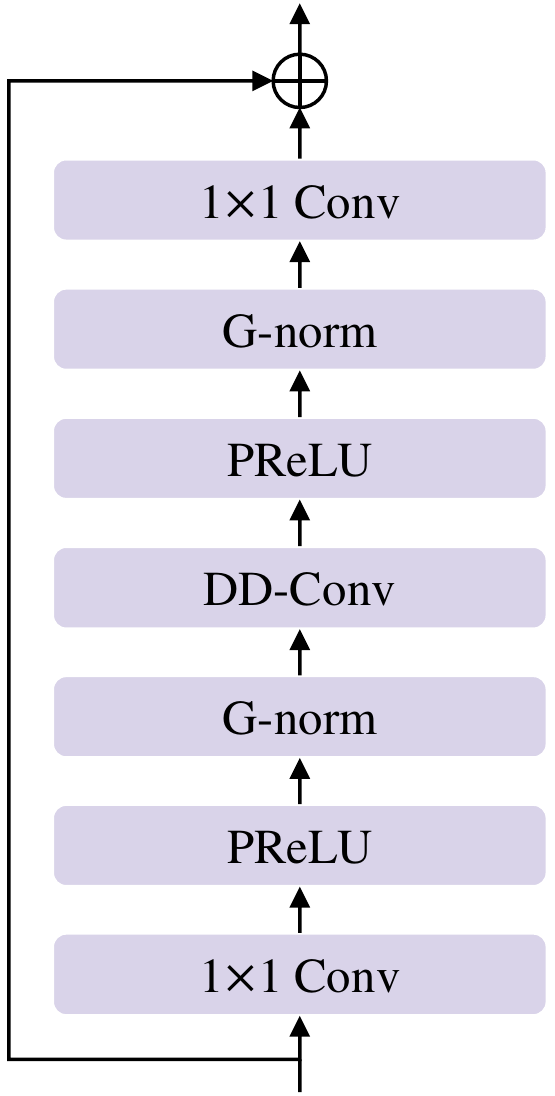}
	\caption{The details of TCN block. The ``DD-Conv" indicates a dilated depth-wise separable convolution. The ``G-norm" is a global layer normalization \cite{luo2019conv}.}
	\label{fig:tcn}
\end{figure}

A lightweight \camodel{} module is proposed to calculate the weights assigned to different frequency bins.
Details of the \camodel{} module are shown in Fig.\ref{arch_b}. 
0Motivated by SpEx \cite{xu2020spex}, we use several parallel 1-D depthwise convolutions with different kernel size, $k_1$(small), $k_2$(middle) and $k_3$(large), to extract the different time-scale features of each frequency bin along the time axis.
The convolutions are followed by average pooling layer and the rectified linear unit (ReLU) activation function to produce the different time-scale features $\mathbf{C}_s$, $\mathbf{C}_m$, $\mathbf{C}_l \in \mathbb{R}^{F}$. 
Then, a fully connected layer is adopted to fuse three features into the fusion feature $\mathbf{C}$.
Eventually, following the operation in \cite{hu2018squeeze}, we stack another two fully connected layers to learn the weights $\mathbf{W}$ from the fusion feature $\mathbf{C}$.

In \model{}, three \camodel{} modules are adopted to weight $\mathbf{X}^{m}$, $\mathbf{X}^{r}$, $\mathbf{X}^{i}$ respectively.
Then the weighted spectrograms $\widetilde{\mathbf{X}}^{m}$, $\widetilde{\mathbf{X}}^{r}$, $\widetilde{\mathbf{X}}^{i} \in \mathbb{R}^{F \times T}$ are fed to full-band extractors $E_{mag}$, $E_{real}$, $E_{imag}$ respectively.

\subsection{Full-band Extractor}
As a model with powerful temporal sequence modeling ability, TCN blocks have been widely used in speech separation \cite{luo2019conv, fan2020end, tzinis2020two}
and target speaker extraction tasks \cite{xu2020spex} recently.
Fig.\ref{fig:tcn} shows the structure of the TCN block, which consists of three main components, namely input $1 \times 1 \ Conv$, depthwise dilated convolution ($DD-Conv$) and output $1 \times 1 \  Conv$.
Parametric ReLU (PReLU) activation function and normalization layers are inserted between adjacent convolutions.
Residual connection is applied to alleviate gradient vanishing problem.
 
In order to further reduce model size and save computation cost, we use stacked TCN blocks to replace LSTMs in full-band model from the original FullSubNet, and name it as full-band extractor.
In detail, a group of stacked TCN blocks is used to replace a layer of LSTM.
Similar to Conv-TasNet \cite{luo2019conv}, we stack the TCN blocks by exponentially increasing the dilation factor in each group to capture the features of speech signals with long-range dependence over the full-band.
As shown in Fig.\ref{arch_b}, for each full-band extractor, there are $M$ groups of $N$ TCN blocks, where $M$ and $N$ are hyper-parameters. 
A fully connected layer is deployed after these stacked TCN blocks.

Full-band extractors extract global context information and output three spectral embeddings $\mathbf{\Psi}^{m}$, $\mathbf{\Psi}^{r}$ and $\mathbf{\Psi}^{i}$ with the same size as their inputs $\widetilde{\mathbf{X}}^{m}$, $\widetilde{\mathbf{X}}^{r}$ and $\widetilde{\mathbf{X}}^{i}$, which are expected to provide complementary information for the sub-band units $\mathbf{\Psi}^{s}$.
Three spectral embeddings and sub-band units 
together serve as the input of $G_{sub}$.

\subsection{Sub-band Model}
For the purpose of learning the frequency-wise signal stationarity and maintaining the stability in model training, as shown in Fig.\ref{arch_b}, the sub-band model $G_{sub}$ uses a structure consisting of two stacked unidirectional LSTM layers and one fully connected layer instead of stacked TCN blocks.

The sub-band model $G_{sub}$ predicts the cIRM based on the weighted noisy sub-band units and the full-band extractors outputs.
In detail, for each frequency $f$, we take a frequency bin vector $\widetilde{\mathbf{X}}^{m}_{f} \in \mathbb{R}^{T}$ and the $2 \times n$ frequency bin vectors adjacent to it in the frequency domain from the weighted magnitude spectrogram $\widetilde{\mathbf{X}}^{m}$ as a sub-band unit $\mathbf{\Psi}^{s}_{f}$:
\begin{equation}
\begin{split}
\mathbf{\Psi}^{s}_{f} =& [\widetilde{\mathbf{X}}^{m}_{f-n}, \cdots ,\widetilde{\mathbf{X}}^{m}_{f}, \cdots , \widetilde{\mathbf{X}}^{m}_{f+n}] \in \mathbb{R}^{(2n+1) \times T}.
\end{split}
\end{equation}
In addition, circular Fourier frequencies are used for boundary frequencies.
The sub-band unit $\mathbf{\Psi}^{s}_{f}$ and the outputs of the full-band extractors, denoted as $\mathbf{\Psi}^{m}_{f}$, $\mathbf{\Psi}^{r}_{f}$, $\mathbf{\Psi}^{i}_{f} \in \mathbb{R}^{T}$ are concatenated with the sub-band unit $\mathbf{\Psi}^{s}_f$ to serve as the input $\mathbf{\Psi}_{f}$ to the $G_{sub}$:
\begin{equation}
\begin{split}
\mathbf{\Psi}_{f} =& [\mathbf{\Psi}^{s}_{f}, \mathbf{\Psi}^{m}_{f}, \mathbf{\Psi}^{r}_{f} , \mathbf{\Psi}^{i}_{f}]^{T}  \in \mathbb{R}^{T \times (2n+4)}.
\end{split}
\end{equation}
We have a total of $F$ such independent input sequences with a dimension of $(T, 2n + 4)$.
Each sequence contains: 1) time-varying signal stationarity and local spectrum patterns in sub-band units;
2) global information and phase information extracted from the magnitude, real and imaginary spectrograms.
They complement each other.
The sub-band model finally makes predictions based on these sequences.

\begin{table*}[!htbp]
    \begin{center}
    \caption{\ninept{The performance in terms of WB-PESQ [MOS], NB-PESQ [MOS], STOI [\%], and SI-SDR [dB] on the DNS Challenge test dataset.}} 
    \label{table:large total compare}
    \scalebox{0.83}{
    \begin{tabular}{cccccccccccc}
    \toprule
        \multirow{2}{*}{Model} & \multirow{2}{*}{Year} &  \multirow{2}{*}{\shortstack{Look Ahead\\(ms)}}  & \multicolumn{4}{c}{With Reverb} & \multicolumn{4}{c}{Without Reverb} \\
    \cmidrule(lr){4-7} \cmidrule(lr){8-11}
    & & &  WB-PESQ & NB-PESQ & STOI & SI-SDR & WB-PESQ & NB-PESQ & STOI & SI-SDR \\
    \midrule
    Noisy & - & - & 1.822 & 2.753 & 86.62 & 9.033 & 1.582 & 2.454 &  91.52 & 9.07 \\
    DCCRN-E\cite{hu2020dccrn} & 2020  & 37.5 & - & 3.077 & - & - & - & 3.266 & - & - \\
    PoCoNet\cite{isik2020poconet} & 2020  & - & 2.832 & - & - & - & 2.748 & - & - & - \\
    DCCRN+\cite{lv2021dccrn+} & 2021  & 10 & - & 3.300 & - & - & - & 3.330 & - & - \\
    TRU-Net\cite{choi2021real} & 2021  & 0 & 2.740 & 3.350 & 91.29 & 14.87 & 2.860 & 3.360 & 96.32 & 17.55 \\
    CTS-Net\cite{li2021two} & 2021  & - & 3.020 & 3.470 & 92.70 & 15.58 & 2.940 & 3.420 & 96.66 & 17.99 \\
    \midrule
    FullSubNet\cite{hao2021fullsubnet} & 2021  & 32  & 3.063 & 3.581 & 92.93 & 16.09 & 2.813 & 3.403 & 96.17 & 17.44 \\
    \model{} & 2021  & 32 & \textbf{3.218} & \textbf{3.666} & \textbf{93.84} & \textbf{16.81} & \textbf{2.982} & \textbf{3.504} & \textbf{96.69} & \textbf{18.34} \\
    \bottomrule
    \end{tabular}}
    \end{center}
\end{table*}
\section{Experiments}
\label{sec:pagestyle}

\subsection{Datasets}
We trained and evaluated \model{} on a subset of the Interspeech 2021 DNS Challenge dataset. 
The clean speech set includes 562.72 hours of clips from 2150 speakers.
The noise dataset includes 181 hours of 60000 clips from 150 classes.
During model training, we use dynamic mixing \cite{hao2021fullsubnet} to simulate speech-noise mixture as noisy speech.
Specifically, before the start of each training epoch, 
75\% of the clean speeches are mixed with the randomly selected room impulse response (RIR) from openSLR26 and openSLR28 \cite{ko2017study} datasets.
After that, the speech-noise mixtures are dynamically generated by mixing clean speeches and noise at a random SNR between -5 and 20 dB.
The DNS Challenge provides a publicly available test dataset consisting of two categories of synthetic clips, namely without and with reverberations. 
Each category has 150 noise clips with a SNR distributed between 0 dB to 20 dB.
We use this test set to evaluate the effectiveness of the model.

\subsection{Training setup and baselines}
We use Hanning window with window length of 32 ms (512-sample) and frame shift of 16 ms to transform the signals to the STFT domain.
Adam optimizer is used with a learning rate of 1e-3.
We set the $\tau$ described above to 2, which is equivalent to exploiting 32 ms information of the next two frames while enhancing the current frame.
For the sub-band units, we set $n=15$ as in \cite{2020subband}.
During training, the input-target sequence pairs are generated as constant-length sequences, with the sequence length set to $T = 192$ frames. 

For a fair comparison, the following two models use the same experimental settings as well as learning target (cIRM).

\textbf{FullSubNet:} 
The model consists of a full-band model and a sub-band model, each containing two layers of stacked LSTMs and one fully connected layer.
The full-band model has 512 hidden units per LSTM layer,
while the sub-band model has 384 hidden units per LSTM layer.

\textbf{\model{}\footnote{The code and examples: \href{https://github.com/thuhcsi/FullSubNet-plus}{https://github.com/thuhcsi/FullSubNet-plus}}:}  The number of channels in the \camodel{} module is 257 and the sizes of the kernels of the parallel 1-D depthwise convolutions are \{3,5,10\} respectively. 
For each full-band extractor, 2 groups of TCN blocks are deployed, each containing 4 TCN blocks with kernel size 3 and dilation rate \{1,2,5,9\}.
The sub-band model contains 384 hidden units per LSTM layer.
The total number of parameters in the model is 8.67 M.


\subsection{Comparison with baseline and SOTA methods}
Table \ref{table:large total compare} shows the performance of each model on the DNS challenge test dataset.
In the table, ``With Reverb" and ``Without Reverb" refer to test sets with and without reverberation respectively.
``Look Ahead" represents the length of future information.
In the last two rows of Table \ref{table:large total compare}, we compare the performance of \model{} with FullSubNet as a baseline.
From the table, we can observe that \model{} has a significant performance improvement over FullSubNet in all metrics.
This improvement shows that the use of the \camodel{} module to help the network focus on more discriminative frequency bands and the use of independent full-band extractor to extract real as well as imaginary spectrograms information are essential for speech enhancement.

\begin{table}[!htbp]
    \begin{center}
    \caption{\ninept{Performance of WB-PESQ and SI-SDR in the ablation study \ref{secsec:ablation1} using the test set without reverberation. Here the backbone is the FullSubNet.}} 
    \label{table:ablation1}
    \scalebox{0.83}{
    \begin{tabular}{ccccccc}
    \toprule
    \multirow{2}{*}{Models} & \multirow{2}{*}{\shortstack{Para\\(M)}} & \multirow{2}{*}{\camodel{}} & \multirow{2}{*}{\shortstack{Full-\\input}} & \multirow{2}{*}{\shortstack{Sub-\\input}} & \multirow{2}{*}{\shortstack{WB-\\PESQ}} & \multirow{2}{*}{\shortstack{SI-\\SDR}} \\
    & & & & & & \\
    \midrule
    FullSubNet$^\spadesuit$ & 5.64 & \XSolidBrush & \XSolidBrush & \XSolidBrush &  2.813 & 17.44 \\
    FullSubNet$^\heartsuit$ & 5.75 & \CheckmarkBold & \CheckmarkBold & \XSolidBrush &  2.900 & 17.71 \\
    FullSubNet$^\clubsuit$ & 5.75 & \CheckmarkBold & \XSolidBrush & \CheckmarkBold &  2.912 & 17.78 \\
    FullSubNet$^\diamondsuit$ & 5.75 & \CheckmarkBold & \CheckmarkBold & \CheckmarkBold & \textbf{2.932} & \textbf{17.90} \\
    \bottomrule
    \end{tabular}}
    \end{center}
\end{table}

\begin{table}[!htbp]
    \begin{center}
    \caption{\ninept{Performance of NB-PESQ, WB-PESQ and SI-SDR in the ablation study \ref{secsec:ablation2} using the test set without reverberation.}} 
    \label{table:ablation2}
    \scalebox{0.83}{
    \begin{tabular}{lccccc}
    \toprule
    \multirow{2}{*}{Models} &  \multirow{2}{*}{\camodel{}} & \multirow{2}{*}{\shortstack{Phase\\-info}} & \multirow{2}{*}{\shortstack{NB-\\PESQ}} & \multirow{2}{*}{\shortstack{WB-\\PESQ}} & \multirow{2}{*}{\shortstack{SI-\\SDR}} \\
    & & & & & \\
    \midrule

    \model{} & \CheckmarkBold & \CheckmarkBold & \textbf{3.504} & \textbf{2.982} & \textbf{18.34} \\    
    $-$ Phase Information & \CheckmarkBold & \XSolidBrush & 3.475 & 2.955 & 17.92 \\
    $-$ \camodel{} Module & \XSolidBrush & \CheckmarkBold & 3.477 & 2.960 & 17.98 \\
    $-$ Both  & \XSolidBrush & \XSolidBrush & 3.453 &  2.900 & 17.73 \\
    \bottomrule
    \end{tabular}}
    \end{center}
\end{table}

Regarding the computational complexity, after using TCN blocks, the parameters of full-band extractor are reduced by 1.64 M.  
Moreover, compared with FullSubNet, the training and inference speed of \model{} on GPU has increased by nearly 18\%.
We also tested it on CPU and \model{} was as much as 80\% faster at inference compared with FullSubNet. 
The processing time for one STFT frame (32 ms) of \model{} on a 2.50 GHz Intel(R) Xeon(R) Platinum 8163 CPU is 18.33 ms, which clearly meets the real-time requirement.


In addition to showing the effectiveness of the improvements for the FullSubNet, we compare \model{} with some current SOTA methods \cite{hu2020dccrn, isik2020poconet, lv2021dccrn+, choi2021real, li2021two} in Table \ref{table:large total compare}.
It can be seen that compared with the latest methods, our proposed \model{} shows superior performance on noise reduction tasks without reverberation and even more prominent performance improvement with reverberation.
This indicates that, as an improvement of the FullSubNet,  \model{} inherits the excellent modeling ability of the sub-band model for reverberation effects described in \cite{hao2021fullsubnet}, and greatly improves the noise reduction ability on this basis.


\subsection{Ablation Study}
\subsubsection{The effect of the \camodel{}}
\label{secsec:ablation1}
We first explore the impact of the \camodel{} module.
Here the backbone is the FullSubNet and the \camodel{} module is used to weight.
We note $^\spadesuit$ as original FullSubNet with no spectrogram weighting,
$^\heartsuit$ as weighting only the input spectrogram of the full-band model,
$^\clubsuit$ as weighting only the input spectrogram of the sub-band model and $^\diamondsuit$ as weighting both the input spectrograms of the full-band model and the sub-band model.
Table \ref{table:ablation1} shows the effect of the model on the test set without reverberation in four cases,
where ``Full-input" and ``Sub-input" refer to weighting the input of the full-band model and weighting the input of the sub-band model, respectively.
As we can see from the table, after the weighting of \camodel{}, both the full-band model and the sub-band model have a gain in performance, and the best performance is obtained when both are used together.
It can be concluded that the \camodel{} module can have a significant effect on the speech enhancement model with only a small increase in the number of parameters.

\subsubsection{The effect of using phase information}
\label{secsec:ablation2}
We then study the effect of adding phase information on the model.
From Table \ref{table:ablation2}, we can observe the results of four variation models on the test set without reverberation,
where ``Phase-info" refers to the use of real and imaginary spectrograms branches to exploit the phase information of noisy speech.
It can be found that the model gains in performance after using the features extracted from the real and imaginary spectrograms branches. 
In addition, this improvement does not conflict with the \camodel{} module, and the combination of both the use of phase information and \camodel{} module gives the best results.
This experimental result confirms that the use of phase information is of great significance to speech enhancement based on sub-band methods, and its combination with \camodel{} module can achieve excellent results.

\section{Conclusions}
\label{sec:conclusions}
In this paper, we have proposed a new single-channel real-time speech enhancement framework named \model{}. 
It adopts the multi-scale time sensitive channel attention (MulCA) to help the network focus on more discriminative frequency bands for noise reduction. 
We also design a full-band extractor which replaces the LSTMs in full-band model from original FullSubNet with stacked TCN blocks to extract suitable feature from all the magnitude, real and imaginary spectrograms. 
The effectiveness of our modules is demonstrated by ablation studies.
We compared FullSubNet+ with other top-ranked methods on the DNS challenge, and the results show that the proposed FullSubNet+ outperforms other nowadays speech enhancement methods.

\textbf{Acknowledgement}: This work is supported by National Natural Science Foundation of China (NSFC) (62076144).

\vfill\pagebreak

\bibliographystyle{IEEEbib}
\bibliography{strings,refs}

\end{document}